# Single-particle theory of persistent spin helices in two-dimensional electron gas: the general approach for quantum wells with different growth direction

Alexander S. Kozulin, Alexander I. Malyshev, and Anton A. Konakov

Department of Theoretical Physics,
National Research Lobachevsky State University of Nizhni Novgorod,
23 Gagarin Avenue, Nizhni Novgorod, 603950, Russia

We present a detailed theoretical investigation of persistent spin helices in two-dimensional electron systems with spin-orbit coupling. For this purpose we consider a single-particle effective mass Hamiltonian with generalized linear-in-*k* spin-orbit coupling term corresponding to a quantum well grown in an arbitrary crystallographic direction, and derive the general condition for the formation of the persistent spin helix. This condition applied for the Hamiltonians describing quantum wells with different growth directions indicates the possibility of existence of the persistent spin helix in a wide class of 2D systems apart from [001] model with equal Rashba and Dresselhaus spin-orbit coupling strengths and the [110] Dresselhaus model. In addition, we employ the translation operator formalism for analytical calculation of space-resolved spin density and visualization of the persistent spin helix patterns.

PACS numbers: 72.25 Dc, 71.70.Ej, 75.70.Tj

# I. INTRODUCTION

Although properties of two-dimensional electron gas (2DEG) with spin-orbit coupling (SOC) have been studied intensively during several decades [1], some of its fascinating features were established relatively recently. In 2006 Bernevig *et al* [2] discovered the new type of SU(2) spin rotation symmetry for two well-known special cases of 2DEG with SOC: [001] model with equal Rashba and Dresselhaus SOC strengths (ReD model) and the Dresselhaus [110] model. They predicted the persistent spin helix (PSH), which is a special spin precession pattern with the precession angle depending only on the net displacement in specific directions (±[110] for ReD model and ±[1$\bar{1}$0] for the Dresselhaus [110] model). Later, the application of the translation operator formalism [3-5] for visualizing the space-resolved spin density allowed to clarify the feature of the PSH state. In 2009, PSHs were experimentally observed in [001]-grown GaAs/AlGaAs quantum wells (QWs) [6], and since then many theoretical [7-13] and experimental [14-19] investigations have been devoted to the studies of different manifestations of the PSH. However, almost all of them primarily focus on two above-mentioned cases and do not cover the other types of 2D electron systems with SOC. The main goal of the present paper is to derive the general condition of the PSH formation in 2DEG contained in a QW grown in an arbitrary crystallographic direction, and outline some specific materials that are good candidates for realization of the PSH state.

This paper is organized as follows. In Sec. II we introduce the generalized form of a single-particle effective mass Hamiltonian with linear-in-*k* SOC and formulate the general condition for the formation of the PSH state in 2D electron systems. Therein, some aspects of the exact SU(2) symmetry responsible for realization of the PSH state are discussed. In Sec. III we apply the general condition obtained in Sec. II for identification of 2D systems in which the PSH patterns may be experimentally observed. In Sec. IV the translation operator formalism is employed for analytical calculation of space-resolved spin density. Section V concludes this work.

# II. THE GENERALIZED SOC-HAMILTONIAN

As stated in the Introduction, our main goal is to derive the general condition of PSH formation in 2DEG formed in a QW grown in an arbitrary crystallographic direction. In general, existence of an extra symmetry connected with the spin degree of freedom leads to conservation of the spin density projection on some specific direction that is characterized by a unit vector $\vec{n}$. In the framework of the single-particle approximation it is expressed as

$$\left[\hat{H},\left(\vec{n}\cdot\hat{\vec{S}}\right)\right]=\hat{0}, \qquad (1)$$

where $\hat{H}$ and $\hat{\vec{S}}$ are the Hamiltonian containing SOC terms and the spin operator, respectively. The condition (1) can be regarded as a starting point of our detailed investigation of the PSHs in a wide class of 2D electron systems with SOC.

Different types of 2D electron systems with SOC are characterized by effective-mass Hamiltonians with various SOC-terms that generally can be derived in accordance with the symmetry of the corresponding system [20]. Motivated by our main goal formulated above, we consider the following generalized form of a single-particle Hamiltonian under the effective mass approximation with linear-in-*k* SOC:

$$\hat{H}=\frac{\hbar^{2}}{2m}\left(\hat{k}_{x}^{2}+\hat{k}_{y}^{2}\right)\hat{\sigma}_{0}+\left(\alpha_{11}\hat{\sigma}_{x}+\alpha_{12}\hat{\sigma}_{y}+\alpha_{13}\hat{\sigma}_{z}\right)\hat{k}_{x}+\left(\alpha_{21}\hat{\sigma}_{x}+\alpha_{22}\hat{\sigma}_{y}+\alpha_{23}\hat{\sigma}_{z}\right)\hat{k}_{y}, \qquad (2)$$

where we use Cartesian coordinates with $z$ axis perpendicular to the plane of 2DEG, $m$ is the electron effective mass, six parameters $\alpha_{ij}$ define the asymmetry induced SOC, $\hat{\sigma}_0$ and $\hat{\sigma}_x$, $\hat{\sigma}_y$, $\hat{\sigma}_z$ are 2×2 unit matrix and the Pauli matrices, respectively. We also assume here the standard relation $\hat{\vec{S}}=\hbar\hat{\vec{\sigma}}/2$.

Calculating the commutator of the Hamiltonian (2) with a linear combination of the Pauli matrices, we find that condition (1) is satisfied if the cross product of two symbolic vectors $\vec{\alpha}_1=\{\alpha_{11},\alpha_{12},\alpha_{13}\}$ and $\vec{\alpha}_2=\{\alpha_{21},\alpha_{22},\alpha_{23}\}$ constructed from the SOC parameters is a null vector:

$$\left[\vec{\alpha}_1\times\vec{\alpha}_2\right]=\vec{0}. \qquad (3)$$

Connection between the relation (3) and the non-Abelian SOC gauge is shown in Appendix A. Condition (3) is satisfied if (a) $\vec{\alpha}_1$ and $\vec{\alpha}_2$ are non-zero and collinear, i.e. holds the relation $\vec{\alpha}_1=q\vec{\alpha}_2$ with a real constant $q$, or (b) either $\vec{\alpha}_1$ or $\vec{\alpha}_2$ is a null vector. In such situations, the operator $\hat{S}_{\vec{\alpha}_1}=\left(\vec{\alpha}_1\cdot\hat{\vec{S}}\right)/|\vec{\alpha}_1|$ of the spin projection on the direction $\vec{\alpha}_1$ commutes with the Hamiltonian (2) and defines an extra conserved physical quantity.

Moreover, in these exceptional symmetric cases, the effective SOC-induced magnetic field with components $\{\alpha_{11}\hat{k}_x+\alpha_{21}\hat{k}_y, \alpha_{12}\hat{k}_x+\alpha_{22}\hat{k}_y, \alpha_{13}\hat{k}_x+\alpha_{23}\hat{k}_y\}$ becomes unidirectional, i.e. its direction does not depend on the components of the wave vector and is defined by the unit vector $\vec{b}=\vec{\alpha}_2/|\vec{\alpha}_2|$. In such situations the SOC-part of the Hamiltonian (2) can be written as

$$\hat{H}_{SO}=\left(q\hat{k}_x+\hat{k}_y\right)\left(\vec{\alpha}_2\cdot\hat{\vec{\sigma}}\right). \qquad (4)$$

In particular, for $q=1$ and $\vec{\alpha}_2 = \{\alpha, -\alpha, 0\}$ the Hamiltonian (4) corresponds to the ReD model, while for $q=0$ and $\vec{\alpha}_2 = \{0, 0, \alpha\}$ we obtain the [110] Dresselhaus model with $y$ axis parallel to $[\bar{1}10]$.

Performing the global spin rotation generated by the operator

$$\hat{U} = \hat{\sigma}_0 \cos \chi/2 - i(n_x \hat{\sigma}_x + n_y \hat{\sigma}_y)\sin \chi/2, \quad (5)$$

where $\vec{n} = [\vec{b} \times \vec{e}_z]$, $\vec{e}_z = \{0, 0, 1\}$, and $\chi$ is the angle between $\vec{e}_z$ and $\vec{b}$, we transform the Hamiltonian (4) to the diagonal form

$$\hat{H}_1 = \frac{\hbar^2}{2m}(\hat{k}_x^2 + \hat{k}_y^2)\hat{\sigma}_0 + |\vec{\alpha}_2|(q\hat{k}_x + \hat{k}_y)\hat{\sigma}_z. \quad (6)$$

Physically, it means that we choose the direction of the effective magnetic field as the spin quantization axis and the quantum states are characterized by the projection of the electron spin on this axis.

The energy spectrum of the Hamiltonian (6)

$$E_\lambda(k_x, k_y) = \frac{\hbar^2}{2m}(k_x^2 + k_y^2) + \lambda|\vec{\alpha}_2|(q k_x + k_y) \quad (7)$$

with $\lambda = \pm 1$ which numerates the energy bands, has an important shifting property:

$$E_{-1}(k_x + Q_x, k_y + Q_y) = E_1(k_x, k_y) \Leftrightarrow E_{-1}(\vec{k} + \vec{Q}) = E_1(\vec{k}), \quad (8)$$

where $\vec{Q} = \frac{2m|\vec{\alpha}_2|}{\hbar^2}\{q, 1\}$ is so-called "magic" vector [2], which physical interpretation in the context of the PSH will be given below. Accordingly, the eigenfunctions of the Hamiltonian (6) corresponding to quantum numbers $\lambda = \pm 1$ have the following simple form:

$$\psi_{\vec{k},1} = \exp(i(\vec{k} \cdot \vec{r}))\begin{Vmatrix}1\\0\end{Vmatrix}, \quad \psi_{\vec{k},-1} = \exp(i(\vec{k} \cdot \vec{r}))\begin{Vmatrix}0\\1\end{Vmatrix}. \quad (9)$$

The ReD and [110] Dresselhaus models being particular cases of the Hamiltonian (6) are characterized by the exact SU(2) symmetry which is introduced in [2] in terms of creation and annihilation operators corresponding to the many-particle problem. Below we demonstrate that the Hamiltonian (6) has the same SU(2) symmetry with generators which are the single-particle counterparts of the generators presented in work [2]. Namely, two single-particle operators

$$\hat{C}_{\vec{Q}}^+ = \frac{\exp(-i(\vec{Q} \cdot \vec{r}))}{2}(\hat{\sigma}_x + i\hat{\sigma}_y) = \exp(-i(\vec{Q} \cdot \vec{r}))\begin{Vmatrix}0 & 1\\0 & 0\end{Vmatrix},$$

$$\hat{C}_{\vec{Q}}^- = \frac{\exp(i(\vec{Q} \cdot \vec{r}))}{2}(\hat{\sigma}_x - i\hat{\sigma}_y) = \exp(i(\vec{Q} \cdot \vec{r}))\begin{Vmatrix}0 & 0\\1 & 0\end{Vmatrix}, \quad (10)$$

that act on the eigenfunctions (9) in the following way

$$\hat{C}_{\vec{Q}}^{+}\psi_{\vec{k},-1} = \psi_{\vec{k}-\vec{Q},1}, \qquad \hat{C}_{\vec{Q}}^{+}\psi_{\vec{k},1} = 0,$$
$$\hat{C}_{\vec{Q}}^{-}\psi_{\vec{k},-1} = 0, \qquad \hat{C}_{\vec{Q}}^{-}\psi_{\vec{k},1} = \psi_{\vec{k}+\vec{Q},-1} \tag{11}$$

together with $\hat{\sigma}_z$ obey the commutation relations for the angular momentum operator

$$\left[\hat{\sigma}_z, \hat{C}_{\vec{Q}}^{+}\right] = 2\hat{C}_{\vec{Q}}^{+}, \quad \left[\hat{\sigma}_z, \hat{C}_{\vec{Q}}^{-}\right] = -2\hat{C}_{\vec{Q}}^{-}, \quad \left[\hat{C}_{\vec{Q}}^{+}, \hat{C}_{\vec{Q}}^{-}\right] = \hat{\sigma}_z. \tag{12}$$

In addition, due to the shifting property (8) $\hat{C}_{\vec{Q}}^{+}$ and $\hat{C}_{\vec{Q}}^{-}$ commute with the Hamiltonian (6):

$$\left[\hat{H}_1, \hat{C}_{\vec{Q}}^{+}\right] = \left(E_1(\vec{k}-\vec{Q}) - E_{-1}(\vec{k})\right)\hat{C}_{\vec{Q}}^{+} = \hat{0}$$
$$\left[\hat{H}_1, \hat{C}_{\vec{Q}}^{-}\right] = \left(E_{-1}(\vec{k}+\vec{Q}) - E_1(\vec{k})\right)\hat{C}_{\vec{Q}}^{-} = \hat{0} \tag{13}$$

It is also obvious that $\left[\hat{H}_1, \hat{\sigma}_z\right] = \hat{0}$. The latter relation together with (13) show that single-particle operators $\hat{C}_{\vec{Q}}^{+}$, $\hat{C}_{\vec{Q}}^{-}$ and $\hat{\sigma}_z$ are generators of the exact SU(2) symmetry responsible for formation of the PSH state. Concrete examples of 2D electron systems where the PSH patterns are expected to appear are given in Sec. III.

## III. WHAT 2D ELECTRON STRUCTURES ARE GOOD CANDIDATES FOR REALIZATION OF THE PSH STATE?

The obtained condition $\left[\vec{\alpha}_1 \times \vec{\alpha}_2\right] = \vec{0}$, in fact, defines the possibility of formation of the PSH patterns in arbitrary 2D electron systems with SOC. Nevertheless, the most attractive structures for experimental realization of the PSH state are expected to be found in specific QWs due to opportunities of the SOC parameters manipulation via (a) external electric field and (b) fitting of QW characteristics (type of material and geometrical parameters) for achieving the above-mentioned relation. In this section, we examine a wide class of semiconductor QWs and outline among them some good candidates for realization of the PSH patterns. In addition, relations between the SOC parameters that should be satisfied in order to achieve the PSH state will be obtained.

We begin the analysis with the zinc-blende type QWs. In such structures the absence of inversion symmetry in the bulk material leads to the Dresselhaus SOC-Hamiltonian:

$$\hat{H}_D = \beta_{ij}\hat{k}_i\hat{\sigma}_j, \tag{14}$$

with six parameters $\beta_{ij}$ ($i = x, y$; $j = x, y, z$). Exact expression of the Dresselhaus term depends on the QW growth direction. The other contribution to the SOC-Hamiltonian is connected with the structure inversion asymmetry and is described by the Rashba term:

$$\hat{H}_R = \alpha_R\left(\hat{k}_x\hat{\sigma}_y - \hat{k}_y\hat{\sigma}_x\right). \tag{15}$$

Hereafter we divide all zinc-blende QWs into symmetric (the corresponding Hamiltonian contains only the Dresselhaus term) and asymmetric (both Dresselhaus and Rashba terms are present) types. SOC-parts of the electron effective mass Hamiltonians corresponding to different QWs are presented in Table I. Application of the criterion (3) for every Hamiltonian from this table allows to conclude whether the PSH state is realized in a specific QW or not.

Firstly, in the mostly studied case of [001] QWs the helices are not formed when a well is symmetric and appear in asymmetric wells only if modules of the Rashba and Dresselhaus SOC strengths are equal (ReD model). This result agrees with previous investigations [2, 3, 6, 7, 15, 16, 18, 19]. Secondly, the PSH structures exist in the symmetric [110] QW for any value of the Dresselhaus SOC strength. This case is known as the Dresselhaus [110] model. In order to guarantee the formation of the PSH it is necessary just to prepare a symmetric well. Therefore, such kind of QWs seems very attractive for realization of the PSH state.

Application of the general condition (3) allows to find all possible 2D systems apart from the ReD and Dresselhaus [110] models which provide the formation of the PSHs. In particular, our calculations show that the PSH patterns can be also realized for some combinations of the SOC parameters in [110]-asymmetric and [113] QWs. Apparently, they are expected to appear in [112] and miscut [001] QWs.

For QWs prepared on [013]-oriented substrates belonging to the trivial point group $C_1$ there are no any restrictions on the relation between the spin and wavevector components arising from the symmetry requirements. Hence, the SOC-Hamiltonian for such systems has the most general form and coincides with the spin-orbit part of the Hamiltonian (2). We pay attention to this case because of the fact that [013]-orientated substrates are used for growth of HgTe-based 2D topological insulators [20, 21].

All cases discussed above are relevant to materials with zinc-blende structure. However, our approach can also be applied for wurzite-type QWs. For example, SOC-part of the Hamiltonians corresponding to GaN and InN QWs grown in [0001] direction reads as $\hat{H} = (\beta+\alpha)(\hat{\sigma}_x\hat{k}_y - \hat{\sigma}_y\hat{k}_x)$, where $x \parallel [11\bar{2}0]$ and $y \parallel [1\bar{1}00]$. In this case two vectors $\vec{\alpha}_1 = \{0, -(\alpha+\beta), 0\}$ and $\vec{\alpha}_2 = \{\alpha+\beta, 0, 0\}$ are not collinear. Therefore, helices are not formed in such type of QWs.

In our analysis we neglected the interface inversion asymmetry, which may yield extra linear-in-*k* terms in the Hamiltonian (2) caused by noninversion symmetric bonding of atoms at heterostructure interface [22-24]. Although such contributions in the SOC have the same form as the Dresselhaus term in [001] QWs made from III-V materials, their inclusion may be important for formation of the PSH state in [110] QWs [25] and essential for SiGe QWs [20, 26]. Therefore, effects of the interface inversion asymmetry should be taken into account in more rigorous investigations of the PSH in the above-mentioned systems.

In summary, our analysis points on possibility of the PSH patterns formation in a wide class of 2D electron systems. In a specific QW the PSH regime can be achieved through manipulation of the Rashba SOC by means of external electric field or (and) adjustment of the Dresselhaus SOC parameters by fitting of the QW width and the other details of its design.

## IV. ANALYTICAL CALCULATION OF SPACE-RESOLVED SPIN DENSITY

With the aim to understand main features of the PSH state more clearly we employ the translation operator formalism developed in [3-5]. Being quite universal this formalism allows to obtain exact expressions for space-resolved spin density in case of arbitrary SOC parameters included in the Hamiltonian (2). Hence, its usage gives us the possibility to visualize the spin density for both SU(2)-symmetric and asymmetric models and reveals some details that distinguish the PSH state from the other ones. We note that results presented below can be also derived under extra assumptions in the framework of the time-dependent approach discussed in [3].

As a basis for further calculations, we solve the Schrödinger equation with the Hamiltonian (2) and find the eigenfunctions

$$\psi_{\vec{k},\lambda}(\vec{r}) = \frac{\exp(i(\vec{k}\cdot\vec{r}))}{\sqrt{|A|^2 + B_\lambda^2}} \begin{Vmatrix} \lambda B_\lambda \exp(i\arg A) \\ |A| \end{Vmatrix} \tag{16}$$

and the energy spectrum of the system

$$E_\lambda(k,\phi) = \frac{\hbar^2 k^2}{2m} + \lambda k \xi(\vec{\alpha}_1, \vec{\alpha}_2, \phi), \tag{17}$$

where $\lambda = \pm 1$ numerates energy bands, $\vec{k} = \{k\cos\phi, k\sin\phi\}$ is the electron momentum, $A = (\alpha_{11} - i\alpha_{12})\cos\phi + (\alpha_{21} - i\alpha_{22})\sin\phi$, $\xi(\vec{\alpha}_1, \vec{\alpha}_2, \phi) = \sqrt{|\vec{\alpha}_1|^2 \cos^2\phi + (\vec{\alpha}_1 \cdot \vec{\alpha}_2)\sin 2\phi + |\vec{\alpha}_2|^2 \sin^2\phi}$, and $B_\lambda = \lambda(\alpha_{13}\cos\phi + \alpha_{23}\sin\phi) + \xi(\vec{\alpha}_1, \vec{\alpha}_2, \phi)$.

Next, we consider an electron with fixed energy and spin direction characterized by polar and azimuthal angles $\theta_s$ and $\varphi_s$, respectively, which is injected inside the plane of 2DEG at point $\vec{r}_0$ and is described by the wave function

$$\psi_0 = \begin{Vmatrix} \exp(-i\varphi_s)\cos(\theta_s/2) \\ \sin(\theta_s/2) \end{Vmatrix}. \tag{18}$$

Then the carrier propagates into the detection point along the direction that forms angle $\phi$ with $x$ axis, while the carrier spin is precessing in the effective SOC-induced magnetic field (see fig. 1). In general, the direction of the effective magnetic field is $\phi$-dependent. Hence, spins of the electrons with unequal $\phi$ are precessing in different ways. Nevertheless, as it was mentioned above, in the symmetric cases (when the condition $[\vec{\alpha}_1 \times \vec{\alpha}_2] = \vec{0}$ holds) the effective magnetic field is uniaxially oriented and the special regime of the spin precession is realized.

For a fixed electron energy and propagation direction ($E$ and $\phi$ are constants) relation (17) for the energy spectrum is an equation which is quadratic in $k$, which has two solutions $k_1$ and $k_{-1}$ corresponding to quantum numbers $\lambda = 1$ and $\lambda = -1$, respectively (see also fig. 2). Consequently, $\psi_0$ can be presented as a linear combination of functions $\psi_{\vec{k}_1,1}$ and $\psi_{\vec{k}_{-1},-1}$:

$$\psi_0 = C_1 \psi_{\vec{k}_1,1} + C_{-1} \psi_{\vec{k}_{-1},-1}, \tag{19}$$

with the expansion coefficients

$$C_\lambda(\vec{r}_0) = \frac{\exp(-i(\vec{k}_\lambda \cdot \vec{r}_0))}{\sqrt{|A|^2 + B_\lambda^2}} \left( \lambda \exp(-i\varphi_s) \cos\left(\frac{\theta_s}{2}\right) B_\lambda \exp(-i \arg A) + |A| \sin\left(\frac{\theta_s}{2}\right) \right).$$

Due to the translation invariance of the system (the Hamiltonian (2) commutes with operators $\hat{k}_x$ and $\hat{k}_y$) the coordinate dependence of the eigenfunctions (16) is presented only in the exponential factor. Therefore, in order to calculate the wave function at point $\vec{r}$ it is sufficient to transform each eigenfunction $\psi_{\vec{k},\lambda}(\vec{r}_0)$ with the corresponding translation operator $\exp[i(\vec{k}_\lambda \cdot (\vec{r} - \vec{r}_0))]$ or, in other words, replace $\vec{r}_0$ by $\vec{r}$ in $\psi_{\vec{k},\lambda}$. Note that the translation operator formalism is not applicable if extra terms that break translation invariance are included in the Hamiltonian.

Assuming for convenience $\vec{r}_0 = \vec{0}$, after some standard algebraic manipulations one can derive the wave function $\psi(\vec{r})$ at the detection point:

$$\psi(\vec{r}) = \exp\left( \frac{i((\vec{k}_1 + \vec{k}_{-1}) \cdot \vec{r})}{2} \right) \sum_{\lambda = \pm 1} \exp(-i\lambda((\vec{k}_{-1} - \vec{k}_1) \cdot \vec{r})/2) C_\lambda(0) \psi_{\vec{k}_\lambda,\lambda}(0). \tag{20}$$

Using the expression (17) for the energy spectrum, we find that $k_{-1} - k_1 = 2m\xi(\vec{\alpha}_1, \vec{\alpha}_2, \phi)/\hbar^2$ and obtain the following relation for the angle of spin precession:

$$\theta(r,\phi) = ((\vec{k}_{-1} - \vec{k}_1) \cdot \vec{r}) = 2m\xi(\vec{\alpha}_1, \vec{\alpha}_2, \phi) r / \hbar^2 \tag{21}$$

which for symmetric cases reads as

$$\theta(r,\phi) = \frac{2m|\vec{\alpha}_2|\sqrt{1+q^2}}{\hbar^2} r \left| \cos\left( \phi - \arctan\frac{1}{q} \right) \right| = |(\vec{Q} \cdot \vec{r})|. \tag{22}$$

It is easy to see that the angle $\theta$ is a function of $r\left|\cos\left(\phi - \arctan\frac{1}{q}\right)\right|$, which is a projection of the electron displacement on the direction of the magic vector (we called it "the PSH axis") that forms the angle $\arctan\frac{1}{q}$ with $Ox$ axis. This means that for different "paths" distinguished by the value of $\phi$, the electron passes unequal distances (with equal projections on the PSH axis), while its spin precessing in the effective magnetic field rotates on the same angle $\theta$ (see fig. 3). In this sense, the PSH axis is the direction of the electron propagation with the largest spin precession rate. Let us note that formula (22) can be viewed as a generalization of previously obtained expressions for spin precession angle for the ReD and [110] Dresselhaus models [3, 7].

Relation (20) allows us to calculate analytically space-resolved spin density (exact expressions are given in Appendix B). Results of these calculations for different values of the SOC parameters are presented in fig. 4. We demonstrate a PSH pattern for the well-studied ReD model in fig. 4a. Another example of a PSH corresponding to the uniaxial effective magnetic field parallel to [111] direction is shown in fig. 4b. In both cases, the electron spin returns exactly to the same orientation after propagating over the distance $L_s = 2\pi/|\vec{Q}|$ along the PSH axis. It is easy to see that in such situations the space-resolved spin density is invariant under translation $L_s \vec{Q}/|\vec{Q}|$.

For comparison, in fig. 4 (c, d) we show the spin density for 2D electron systems where SU(2) symmetry is absent. Taking unequal the Rashba and Dresselhaus SOC parameters in [001]-grown QW, we break the SU(2) symmetry and the helix disappears (fig. 4c). Similarly, in an arbitrary symmetric case it is sufficient to change only one SOC parameter $\alpha_{11}$ for deviation from the condition of the PSH formation (fig. 4d). For both these situations $[\vec{\alpha}_1 \times \vec{\alpha}_2] \neq \vec{0}$ and, therefore, the structure of the space-resolved spin density differs from the PSH patterns corresponding to the SU(2)-symmetric cases.

## V. CONCLUSIONS

In conclusion, we have theoretically studied the spin precession in two-dimensional electron systems with spin-orbit coupling and have derived the general condition of the persistent spin helix state formation. This condition applied for the Hamiltonians describing quantum wells with different growth directions indicates on the possibility of existence of the PSH patterns in a wide class of 2D systems including well-studied [001] model with equal the Rashba and Dresselhaus

SOC strengths and the [110] Dresselhaus model. The latter statement requires an experimental verification and we hope that appropriate studies will be realized soon.

**ACKNOWLEDGMENTS**

The authors are grateful to G.M. Maksimova for valuable discussions. We also would like to thank D.V. Khomitsky for technical assistance. The work was in part supported by RFBR (grants no. 15-42-02254, 16-07-01102, 16-32-00683, 16-32-00712 and 16-57-51045) and by the Russian Ministry of Education and Science (project no. 3.285.2014/K).

**APPENDIX A: THE GENERAL CONDITION FOR THE PSH FORMATION IN TERMS OF THE NON-ABELIAN SPIN-ORBIT GAUGE**

In this Appendix, we formulate the general condition for realization the PSH state in terms of the non-Abelian SO gauge. In [7, 27] it was shown that the Rashba and Dresselhaus SOC in two dimensions can be regarded as a Yang-Mills non-Abelian gauge field. Following the approach developed in [7], we introduce the SOC gauge

$$\hat{\vec{\gamma}} = \{\hat{\gamma}_x, \hat{\gamma}_y\} = \frac{m}{\hbar}\{(\vec{\alpha}_1 \cdot \hat{\vec{\sigma}}), (\vec{\alpha}_2 \cdot \hat{\vec{\sigma}})\}, \tag{A1}$$

and rewrite the Hamiltonian (2) in the form

$$\hat{H} = \frac{1}{2m}\left((\hat{p}_x\hat{\sigma}_0 + \hat{\gamma}_x)^2 + (\hat{p}_y\hat{\sigma}_0 + \hat{\gamma}_y)^2\right) - V\hat{\sigma}_0, \tag{A2}$$

with the constant potential $V = (|\vec{\alpha}_1|^2 + |\vec{\alpha}_2|^2)/2\hbar^2$. In general, the components of the SO gauge do not commute,

$$[\hat{\gamma}_x, \hat{\gamma}_y] = \frac{2im^2}{\hbar^2}(\hat{\vec{\sigma}} \cdot [\vec{\alpha}_1 \times \vec{\alpha}_2]). \tag{A3}$$

However, in the symmetric cases, when $[\vec{\alpha}_1 \times \vec{\alpha}_2] = \vec{0}$, the commutator (A3) identically vanishes. The latter means that two relations, $[\vec{\alpha}_1 \times \vec{\alpha}_2] = \vec{0}$ and $[\hat{\gamma}_x, \hat{\gamma}_y] = \hat{0}$, are equivalent mathematical formulations of the PSH state formation condition in the framework of the generalized model with linear-in-*k* SOC term.

# APPENDIX B: AN EXACT EXPRESSION FOR SPACE-RESOLVED SPIN DENSITY

Here, we derive the exact expressions for the space-resolved spin density obtained in the framework of the translation operator formalism. The known relation (21) for the spin precession angle allows us to write the explicit expression for the wave function at the detection point $\vec{r}$:

$$\psi(\vec{r}) = \exp\left(\frac{i\left((\vec{k}_1 + \vec{k}_{-1})\cdot\vec{r}\right)}{2}\right) \left\| \begin{array}{l} \exp(-i\varphi_s)\cos\left(\dfrac{\theta_s}{2}\right)f_1(\theta) + A\sin\left(\dfrac{\theta_s}{2}\right)f_2(\theta) \\ A^* \exp(-i\varphi_s)\cos\left(\dfrac{\theta_s}{2}\right)f_2(\theta) + |A|^2 \sin\left(\dfrac{\theta_s}{2}\right)f_3(\theta) \end{array} \right\|, \quad (B1)$$

where we introduce for brevity three functions

$$f_1(\theta) = \sum_{\lambda=\pm 1}\frac{B_\lambda^2}{|A|^2 + B_\lambda^2}\exp\left(-\frac{i\lambda\theta}{2}\right), \quad f_2(\theta) = \sum_{\lambda=\pm 1}\frac{\lambda B_\lambda}{|A|^2 + B_\lambda^2}\exp\left(-\frac{i\lambda\theta}{2}\right),$$

$$f_3(\theta) = \sum_{\lambda=\pm 1}\frac{1}{|A|^2 + B_\lambda^2}\exp\left(-\frac{i\lambda\theta}{2}\right),$$

and analytically calculate space-resolved spin density as $\langle\vec{S}\rangle = \psi^+(\vec{r})\hat{\vec{S}}\psi(\vec{r})$:

$$\left\| \begin{array}{l} \langle S_x \rangle \\ \langle S_y \rangle \\ \langle S_z \rangle \end{array} \right\| = \frac{\hbar}{2} \left\| \begin{array}{l} \operatorname{Re} F_1(\theta) \\ -\operatorname{Im} F_1(\theta) \\ F_2(\theta) \end{array} \right\|, \quad (B2)$$

with functions

$$F_1(\theta) = 2A\left(\cos^2\left(\frac{\theta_s}{2}\right)f_1(\theta)f_2^*(\theta) + |A|^2\sin^2\frac{\theta_s}{2}f_2(\theta)f_3^*(\theta)\right) +$$

$$\left(|A|^2 \exp(-i\varphi_s)f_1(\theta)f_3^*(\theta) + A^2|f_2|^2\exp(i\varphi_s)\right)\sin(\theta_s)$$

and

$$F_2(\theta) = \cos^2\left(\frac{\theta_s}{2}\right)\left(|f_1(\theta)|^2 - |A|^2|f_2(\theta)|^2\right) + |A|^2\sin^2\left(\frac{\theta_s}{2}\right)\left(|f_2(\theta)|^2 - |A|^2|f_3(\theta)|^2\right) +$$

$$\sin(\theta_s)\operatorname{Re}\left(A\exp(i\varphi_s)\left(f_1^*(\theta)f_2(\theta) - |A|^2 f_2^*(\theta)f_3(\theta)\right)\right).$$

Relation (B2) is the final expression for the space-resolved spin density that allows us to calculate its components at an arbitrary point of the 2DEG plane.

# REFERENCES

1. R. Winkler, *Spin-Orbit Coupling Effects in Two-Dimensional Electron and Hole Systems* (Springer-Verlag, Berlin, Heildelberg, 2003).
2. B.A. Bernevig, J. Orenstein, S.-C. Zhang, Phys. Rev. Lett. **97**, 236601 (2006).
3. M.-H. Liu, K.-W. Chen, S.-H. Chen, and C.-R. Chang, Phys. Rev. B **74**, 235322 (2006).
4. M.-H. Liu, C.-R. Chang, Journal of Magnetism and Magnetic Materials **304**, 97 (2006).
5. M.-H. Liu and C.-R. Chang, Phys. Rev. B **74**, 195314 (2006).
6. J.D. Koralek *et al*, Nature (London) **458**, 610 (2009).
7. S.-H. Chen and C.-R. Chang, Phys. Rev. B **77**, 045324 (2008).
8. M.C. Lüffe, J. Kailasvuori, and T.S. Nunner, Phys. Rev. B **84**, 075326 (2011).
9. M.C. Lüffe, J. Danon, and T.S. Nunner, Phys. Rev. B **87**, 125416 (2013).
10. V.A. Slipko, I. Savran, and Y.V. Pershin, Phys. Rev. B **83**, 193302 (2011).
11. M.A.U. Absor, F. Ishii, H. Kotaka, and M. Saito, Applied Physics Express **8**, 073006 (2015).
12. Xin Liu and Jairo Sinova, Phys. Rev. B **86**, 174301 (2012).
13. Vincent E. Sacksteder, IV and B.A. Bernevig, Phys. Rev. B **89**, 161307 (R) (2014).
14. M. Kohda *et al*, Phys. Rev. B **86**, 081306(R) (2012).
15. M.P Walser, C. Reichl, W. Wegscheider, and G. Salis, Nature Physics **8**, 757 (2012).
16. P. Altmann *et al*, Phys. Rev. B **90**, 201306(R) (2014).
17. A. Sasaki *et al*, Nature Nanotechnology **9**, 703 (2014)
18. C. Schönhuber *et al*, Phys. Rev. B **89**, 085406 (2014).
19. G. Salis *et al*, Phys. Rev. B **89**, 045304 (2014).
20. S.D. Ganichev and L.E. Golub, Phys. Status Solidi B **251**, 1801 (2014).
21. K.-M. Dantscher *et al*, Phys. Rev. B **92**, 165314 (2015).
22. O. Krebs and P. Voisin, Phys. Rev. Lett. **77**, 1829 (1997).
23. O. Krebs *et al*, Semicond. Sci. Technol. **12**, 938 (1997).
24. L. Vervoort and P. Voisin, Phys. Rev. B **56**, 12744 (1997).
25. M.O. Nestoklon, S.A. Tarasenko, J.-M. Jancu, and P. Voisin, Phys. Rev. B **85**, 205307 (2012).
26. M.O. Nestoklon, L.E. Golub, and E.L. Ivchenko, Phys. Rev B **73**, 235334 (2006).
27. N. Hatano, R. Shirasaki, and H. Nakamura, Phys. Rev. A **75**, 032107 (2007).

FIGURES

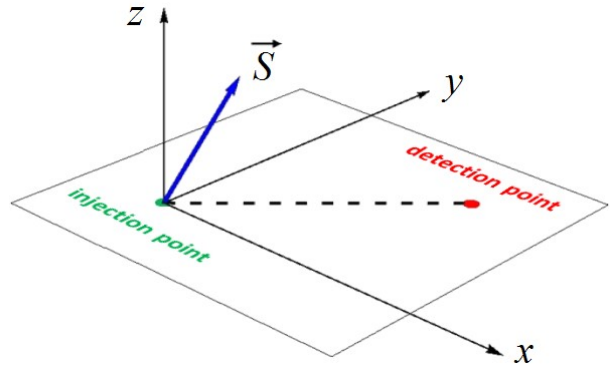

**Figure 1.** The schematic illustration of an electron "path" in the plane of 2DEG. A particle with an initial spin (the blue arrow) is placed in the injection point (the green spot) and then propagates into the detection point (the red spot) with polar coordinates $r$ and $\phi$, while its spin is precessing in the effective magnetic field.

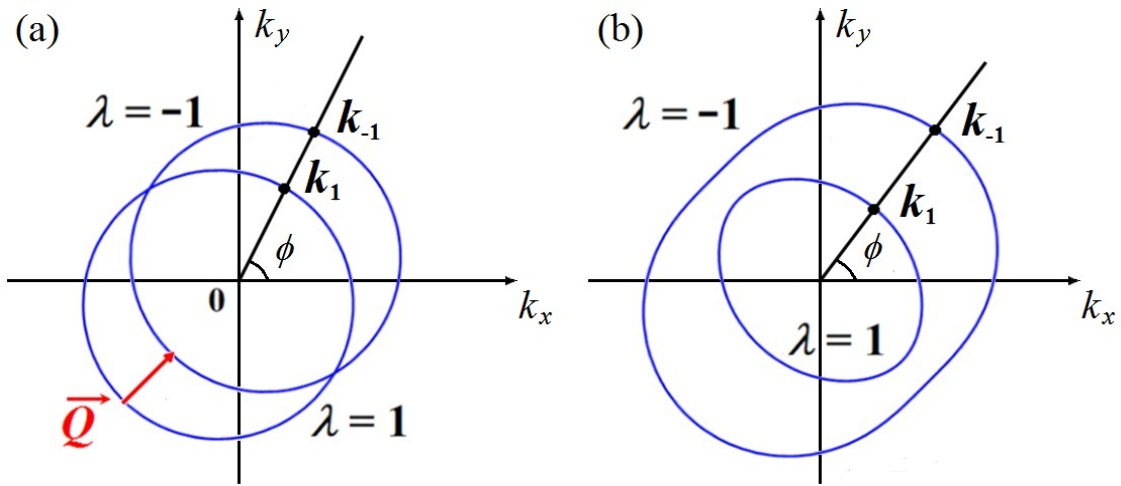

**Figure 2.** Fermi surfaces for (a) symmetric and (b) asymmetric cases. For fixed electron energy $E$ and propagation direction $\phi$ the wave function $\psi_0$ is a linear combination of functions $\psi_{\vec{k}_1,1}$ and $\psi_{\vec{k}_{-1},-1}$. In symmetric cases the Fermi surfaces are two circles shifted by the magic vector $\vec{Q}$.

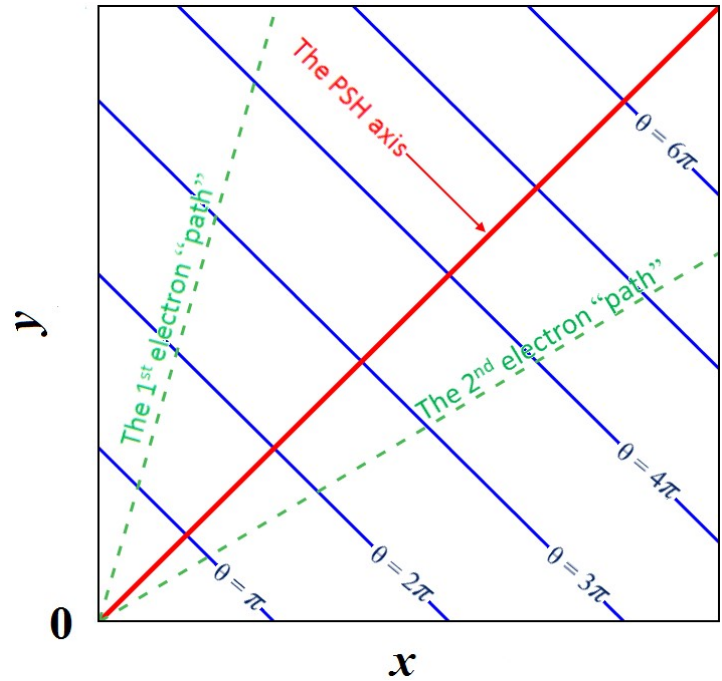

**Figure 3.** The PSH axis as the direction with the largest spin precession rate. In general, electrons pass unequal distances moving along different "paths" (the green dashed lines), while their spins are rotating on the same angle. Nevertheless, in a symmetric case when $[\vec{\alpha}_1 \times \vec{\alpha}_2] = \vec{0}$ the curves of constant spin precession angle $\theta$ (the blue lines) are straight lines and the precession angle depends only on the net displacement in the direction of the PSH axis (the red line).

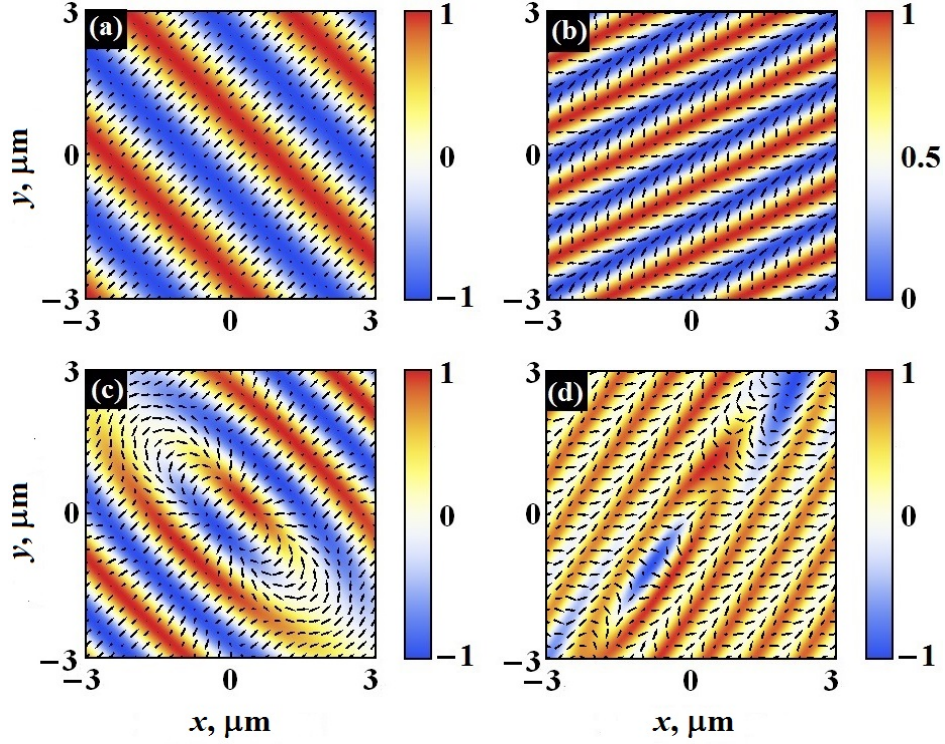

**Figure 4.** Space-resolved spin densities for different SOC parameters: (a) $\vec{\alpha}_1 = \vec{\alpha}_2 = \{\alpha_0, -\alpha_0, 0\}$ (ReD model); (b) $\vec{\alpha}_1 = -\vec{\alpha}_2 = \{2\alpha_0, 2\alpha_0, 2\alpha_0\}$ (a PSH pattern for the out-of-plane uniaxial effective magnetic field); (c) $\vec{\alpha}_1 = \{\alpha_0, -2\alpha_0, 0\}$, $\vec{\alpha}_2 = \{2\alpha_0, -\alpha_0, 0\}$ ([001]-model with unequal Rashba and Dresselhaus SOC strengths); (d) All SOC parameters are the same as for (b), except for $\alpha_{11} = 4\alpha_0$ (an arbitrary situation corresponding to a 2D system with absence of SU(2) symmetry). For all cases $m = 0.02 m_0$, where $m_0$ stands for the free electron mass and $\alpha_0 = 2.5$ meV·nm. The injection is assumed at point $(0,0)$ with $\varphi_s = \pi/4$ and $\theta_s = \pi/2$. Arrows express projection of $\langle \vec{S} \rangle$ onto the 2DEG plane, bars calibrate $\langle S_z \rangle$.

**TABLE I.** Formation of persistent spin helices in quantum wells with different growth direction. All SOC-Hamiltonians are taken from the review [20] and correspond to materials with zinc-blende structure.

| № | Grown direction (z-axis) and orientation of x- and y- axes | Type | SOC-Hamiltonian | $\vec{\alpha}_1$ and $\vec{\alpha}_2$ | Existence of the PSH | Relation between the SOC parameters when the PSH state is realized |
|---|---|---|---|---|---|---|
| 1 | [001], $x \parallel [100]$, $y \parallel [010]$ | symmetric | $\hat{H} = \beta(\hat{k}_x \hat{\sigma}_x - \hat{k}_y \hat{\sigma}_y)$ | $\vec{\alpha}_1 = \{\beta, 0, 0\}$, $\vec{\alpha}_2 = \{0, -\beta, 0\}$ | No | |
| | | asymmetric | $\hat{H} = \beta(\hat{k}_x \hat{\sigma}_x - \hat{k}_y \hat{\sigma}_y) + \alpha(\hat{k}_y \hat{\sigma}_x - \hat{k}_x \hat{\sigma}_y)$ | $\vec{\alpha}_1 = \{\beta, -\alpha, 0\}$, $\vec{\alpha}_2 = \{\alpha, -\beta, 0\}$ | Yes | $|\alpha| = |\beta|$ |
| 2 | [110], $x \parallel [\bar{1}10]$, $y \parallel [001]$ | symmetric | $\hat{H} = \beta \hat{k}_x \hat{\sigma}_z$ | $\vec{\alpha}_1 = \{0, 0, \beta\}$, $\vec{\alpha}_2 = \{0, 0, 0\}$ | Yes | $\beta \neq 0$ |
| | | asymmetric | $\hat{H} = (\beta_1 + \alpha)\hat{k}_y \hat{\sigma}_x + (\beta_2 - \alpha)\hat{\sigma}_y \hat{k}_x + \beta_3 \hat{k}_x \hat{\sigma}_z$ | $\vec{\alpha}_1 = \{0, \beta_2 - \alpha, \beta_3\}$, $\vec{\alpha}_2 = \{\beta_1 + \alpha, 0, 0\}$ | Yes | $\beta_1 + \alpha = 0$ or $(\beta_2 - \alpha)^2 + \beta_3^2 = 0$ |
| 3 | [111], $x \parallel [11\bar{2}]$, $y \parallel [\bar{1}10]$ | symmetric | $\hat{H} = \beta(\hat{k}_y \hat{\sigma}_x - \hat{k}_x \hat{\sigma}_y)$ | $\vec{\alpha}_1 = \{0, -\beta, 0\}$, $\vec{\alpha}_2 = \{\beta, 0, 0\}$ | No | |
| | | asymmetric | $\hat{H} = (\alpha + \beta)(\hat{k}_y \hat{\sigma}_x - \hat{k}_x \hat{\sigma}_y)$ | $\vec{\alpha}_1 = \{0, -(\alpha + \beta), 0\}$, $\vec{\alpha}_2 = \{\alpha + \beta, 0, 0\}$ | No | |
| 4 | [113], $x \parallel [1\bar{1}0]$, $y \parallel [33\bar{2}]$ | symmetric | $\hat{H} = \beta_1 \hat{\sigma}_x \hat{k}_y + \beta_2 \hat{\sigma}_y \hat{k}_x + \beta_3 \hat{\sigma}_z \hat{k}_x$ | $\vec{\alpha}_1 = \{0, \beta_2, \beta_3\}$, $\vec{\alpha}_2 = \{\beta_1, 0, 0\}$ | Yes | $\beta_1 = 0$ or $\beta_2 = \beta_3 = 0$ |
| | | asymmetric | $\hat{H} = (\beta_1 + \alpha)\hat{k}_y \hat{\sigma}_x + (\beta_2 - \alpha)\hat{\sigma}_y \hat{k}_x + \beta_3 \hat{k}_x \hat{\sigma}_z$ | $\vec{\alpha}_1 = \{0, \beta_2 - \alpha, \beta_3\}$, $\vec{\alpha}_2 = \{\beta_1 + \alpha, 0, 0\}$ | Yes | $\beta_1 + \alpha = 0$ or $(\beta_2 - \alpha)^2 + \beta_3^2 = 0$ |
| 5 | [013], x- and y-axes are orthogonal to z-axis | symmetric | $\hat{H} = \sum_{lm} \beta_{lm} \hat{\sigma}_l \hat{k}_m$, $l = \overline{1,3}$, $m = \overline{1,2}$ | $\vec{\alpha}_1 = \{\beta_{11}, \beta_{12}, \beta_{13}\}$, $\vec{\alpha}_2 = \{\beta_{21}, \beta_{22}, \beta_{23}\}$ | Yes | $\vec{\alpha}_1$ and $\vec{\alpha}_2$ are collinear |
| | | asymmetric | $\hat{H} = \sum_{lm} \beta_{lm} \hat{\sigma}_l \hat{k}_m + \alpha(\hat{k}_y \hat{\sigma}_x - \hat{k}_x \hat{\sigma}_y)$ | $\vec{\alpha}_1 = \{\beta_{11}, \beta_{12} - \alpha, \beta_{13}\}$, $\vec{\alpha}_2 = \{\beta_{21} + \alpha, \beta_{22}, \beta_{23}\}$ | Yes | |